\documentclass[aps,pra,twocolumn,showpacs,preprintnumbers,superscriptaddress]{revtex4}
\usepackage{amsmath}
\usepackage{amsthm}
\usepackage{graphicx}
\usepackage{dcolumn}
\usepackage{bm}
\usepackage{appendix}
\usepackage{color}
\definecolor{mycolor}{rgb}{0,0.5,1}

\newtheorem{theorem}{Theorem}
\newtheorem{lemma}{Lemma}

\begin{document}

\title{Security proof of differential phase shift quantum key distribution
in the noiseless case}

\begin{abstract}
Differential phase shift quantum key distribution systems have a high
potential for achieving high speed key generation. However, its
unconditional security proof is still missing, even though it has been
proposed for many years. Here, we prove its security against collective
attacks with a weak coherent light source in the noiseless case (i.e. no bit
error). The only assumptions are that quantum theory is correct, the devices
are perfect and trusted and the key size is infinite. Our proof works on
threshold detectors. We compute the lower bound of the secret key generation
rate using the information-theoretical security proof method. Our final
result shows that the lower bound of the secret key generation rate per
pulse is linearly proportional to the channel transmission probability if
Bob's detection counts obey the binomial distribution.
\end{abstract}

\author{Yi-Bo Zhao}
\affiliation{Key Lab of Quantum Information, University of Science and Technology of
China, (CAS), Hefei, Anhui 230026, China}
\author{Chi-Hang Fred Fung}
\affiliation{Department of Physics and Center of Computational and Theoretical Physics,
University of Hong Kong, Pokfulam Road, Hong Kong, China}
\author{Zheng-Fu Han}
\affiliation{Key Lab of Quantum Information, University of Science and Technology of
China, (CAS), Hefei, Anhui 230026, China}
\author{Guang-Can Guo}
\affiliation{Key Lab of Quantum Information, University of Science and Technology of
China, (CAS), Hefei, Anhui 230026, China}

 \maketitle

\section{Introduction}

Quantum key distribution (QKD) allows
two distant parties to share secret keys that are unconditionally secure.
Until now, there have been several kinds of QKD protocols. The traditional
BB84 protocol is one based on qubits, in which Alice sends Bob a sequence of
qubits to establish a secret key \cite{BB84}. On the other hand, there are
other non-qubit-based protocols,
such as the continuous variable QKD scheme \cite{CVQKD,Heid,zhaoyb}.
In these two protocols, regardless of whether qubit states or
continuous states are used, each state (or state pair) received by
Bob directly gives rise to one bit value. In contrast, in
differential phase shift (DPS) QKD, information is
encoded in the difference between each two adjacent quantum states \cite%
{Inoue,Inoue2,Yamamoto}.
Of the above 
protocols, the last one is designed to achieve high speed communication.
In Ref. \cite{Yammamoto experiment}, Diamanti, et al. have realized
DPSQKD with a modulation frequency of 1 GHz.

DPSQKD is well suited for coherent-state sources as information is
encoded in the relative phases of coherent states. Coherent-state
sources can also be used with other protocols, for example, the BB84
protocol with phase-randomized coherent states~\cite{Gottesman2004}
whose performance is substantially improved by the decoy-state
method~\cite{Hwang2003,Lo2005,Ma2005b,Wang2005a,Wang2005b,Harrington2005,Zhao2006,Zhao2006b},
the BB84 protocol with phase-non-randomized coherent
states~\cite{Lo2007}, and the B92 protocol with strong reference
pulses~\cite{Bennett1992,Koashi2004-B92,Tamaki2006-B92} (see also
\cite{Tamaki2008}).

The unconditional security of the BB84 protocol is well discussed
(e.g.,
\cite{Mayers2001,Biham2000,Lo1999,Shor2000,Gottesman2004,Inamori2005}).
For the CVQKD protocols, its security against collective attack is
also well discussed \cite{Heid,zhaoyb,Garcia}. However, for DPSQKD,
we only know that it is secure against several specific attacks,
e.g., the beam splitting attack and the intercept and resending
attack \cite{Yammamoto experiment,individual attack}. Until now, we
do not know whether it is secure against any quantum Eve even in the
noiseless case. On the other hand, specific attacks on DPSQKD have
been proposed to evaluate upper bounds on the secret key generation
rates of
DPSQKD 
\cite{Sequential attack,Sequential attack 2,Upper bound,Hipolito,Sequential
attack3}.

In this paper, we prove the security of DPSQKD against collective attacks
\cite{explanation of collective} with a weak coherent light source in the
noiseless case. The security in this case follows from a key result that we
will prove in this paper, namely that Eve's state is independent of the
positions of Bob's detected signals. This result makes sense since the fact
that there is no bit error restricts what Eve can do to Bob's signals. In
particular, she has to ensure that Bob receives all signals with equal
intensities, since signals with different intensities will result in bit
errors with non-zero probability.
Our final result on the lower bound on the key generation rate is a function
of the estimated parameters of the channel (see Eq.~
\eqref{final secret key
rate}). In order to understand this result further, we compute this bound by
considering a channel that produces a binomial distribution in Bob's
detection statistics. Specifically, we show that the lower bound of the
secret key generation rate per pulse is linearly proportional to the channel
transmission probability (see Eq.~\eqref{key rate per pulse}).

The only assumptions used in the proof are that quantum theory is
correct, the key size is infinite and the devices are perfect and
trusted. Our proof works on threshold detectors, which are the
detectors commonly used in practice. Furthermore, we do not require
quantum non-demolition (QND) measurements.
Even though we consider collective attacks in this
paper,
we speculate that
it is possible for us to
extend our security proof to the most
general attack, namely the coherent attack. An intuitive justification tells
us that Eve cannot get more information from coherent attacks than that from
collective attacks as the key size goes to infinite \cite{Review,Renner}.
For the finite dimension case, there is a exponential de Finetti theorem
that can strictly give this result \cite{de Finitte,Renner2}. However, for
DPSQKD, the states sent by Alice are weak coherent states and thus in theory
the dimension of Bob's received states may be infinite. Thus, the current de
Finetti theorem cannot be directly applied. On the other hand, there are
three potential
ways to solve this problem. Since in practice the states sent by Alice are
weak coherent states and the probability that Bob gets a large photon number
is extremely small, it is possible to prove that Alice and Bob's states can
be well approximated by states that have a finite support. Then the current
de Finetti theorem can be applied. The second way is to extend the current
exponential de Finetti theorem to the infinite dimensional case.
We can see a hope of this in Ref. \cite{Toner}. The third way is to extend
the current de Finetti theorem to the case with finite number of measurement
results, since we know that in DPSQKD, Bob's measurement results are finite.
The work in Ref. \cite{Toner} also shows that this way may be viable.

We note that there is also a recent proof \cite{Wen2008} on the DPS
protocol. Their proof assumes a single-photon source and requires
QND measurements, whereas our proof allows a more general weak
coherent light source and does not require QND measurements. On the
other hand, their proof can handle the noisy case and applies to the
most general attack, whereas ours can only handle noiseless
collective attacks.

In the following analysis, we map the traditional {DPS} protocol into a
big-state protocol and give the security proof for this big-state protocol.
In doing so, we prove a key result of our paper, which is that Eve's state
is independent of the positions of Bob's detected signals. With this result
and some properties of the mutual information, we can upper bound Eve's
information about Alice's bit string. Finally, we evaluate the key rate
assuming a typical setting in which the detection statistics follows the
binomial distribution.
The security proof method we employ is the information-theoretical one
by Renner, et al. \cite{Renner,Renner2}.

\section{Equivalent Protocols}

\subsection{Protocol 1 - original protocol}

\begin{description}
\item[Quantum Phase:]

1. Alice sends a sequence of coherent states, each with amplitude $\alpha$,
but with a randomly selected phase, $|\alpha \rangle $ or $|-\alpha \rangle $%
, to Bob. Then she records each state with a binary variable $x_{i}$, by
setting $x_{i}=0$ if the $i$-th state is $|-\alpha \rangle $ and $x_{i}=1$
if the $i$-th state is $|\alpha \rangle $.

2. Using the Mach-Zehnder (M-Z) interferometers shown in Fig. 1, Bob
measures the phase difference between every two adjacent states. Bob stores
his measurement result into binary variables $y_{i}$ and $z_{i}$. While
measuring the phase difference of the $i$-th and $(i-1)$-th state, if Bob
gets a photon count, he sets $z_{i}=1$ and if not, he sets $z_{i}=0$. Also,
if $i=kN+1$ ($k=0,1,2,3,...$), Bob sets $z_{i}=0.$\ If $z_{i}=1$\ and the
measurement results indicates that the phase difference is zero, he sets $%
y_{i}=0$; otherwise, if it indicates a non-zero phase difference, he sets $%
y_{i}=1$. If $z_{i}=0$, Bob sets $y_{i}=0$.

\item[Classical Phase:]

3. After Bob receives each set of $N$ states, he announces all $z_{i}$'s for
these $N$ states.

4. Alice generates another variable $l_{i}=x_{i}\oplus x_{i-1}$ for $i>1$.

5. After many rounds of such communications, Alice and Bob randomly publish
some of $l_{i}$ and $y_{i}$ corresponding to $z_{i}=1$ to test the bit error
rate (BER) between them.

6. Alice and Bob generate new binary variables $u_{i}^{A}$ and $u_{i}^{B}$
respectively. Alice sets $u_{i}^{A}=l_{i}\cdot z_{i}$. Bob sets $%
u_{i}^{B}=y_{i}\cdot z_{i}$.

7. Alice and Bob estimate the mutual information between the binary strings $%
\vec{u}^{A}$ and $\vec{u}^{B}$ conditioned on Bob's announcement.

8. Alice announces the error correction information of binary string $\vec{u}%
^{A}$ and Bob uses it to reconcile his string $\vec{u}^{B}$ to the
corresponding string $\vec{u}^{A}$.

9. Alice and Bob perform privacy amplification on their common binary string
$\vec{u}^{A}$ to generate the final secret key.
\end{description}

Instead of steps 6-9, Alice and Bob can simply discard the $l_{i}$\ and $%
y_{i}$ that correspond to $z_{i}=0$, and perform error correction and
privacy amplification on the remaining sifted bits. This protocol is then
equivalent to 
the original DPSQKD protocol \cite{Inoue,Yamamoto}. In this protocol, Bob
measures the phase difference between every two adjacent pulses. Thus, it
may not be a good idea to try to map them into single bits and then discuss
the security. Here, our basic idea is to regard $N$ states as one big state
and to discuss the security of this big state. In the following, we will
introduce three protocols that map the above protocol into a big state
protocol, in which Alice sends Bob a big state and Bob measures it with $N$
equipments. The security of these three protocols are weaker than the above
one. Thus, if the security of these inferior protocols are proved, then the
security of the above protocol is proved.

\begin{figure}[tbp]
\includegraphics[width=8cm]{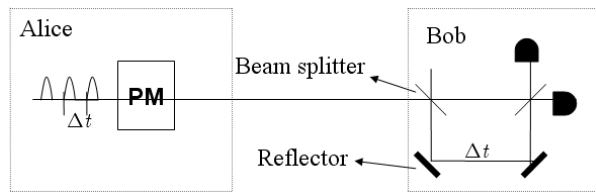}
\caption{Illustration of the protocol 1, where the PM denotes the phase
modulator}
\label{fig1}
\end{figure}

\subsection{Protocol 2}

\begin{figure}[tbp]
\includegraphics[width=8cm]{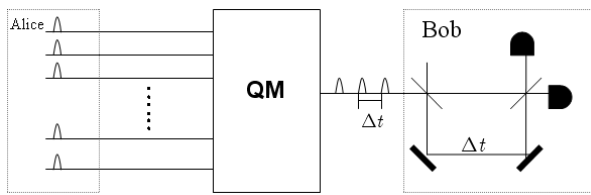}
\caption{Illustration of Protocol 2, where the QM system is a quantum memory
that stores the $N$ received states and sends out them one by one.}
\label{fig2}
\end{figure}

In step 1, according to the binary string $\vec{x}%
=(x_{kN+1},x_{kN+2},...,x_{(k+1)N})$, Alice generates a state $|\Psi _{\vec{x%
}}^{N}\rangle ={\bigotimes\limits_{i=kN+1}^{(k+1)N}}|(-1)^{x_{i}+1}\alpha
\rangle $ and sends it to Bob through $N$ fibers. In the quantum channel,
there is a quantum memory (QM) system that separates $|\Psi _{\vec{x}%
}^{N}\rangle $ into an $N$-state sequence and sends each state to Bob one by
one through one fiber (see Fig. \ref{fig2}). Steps 2-9 remain the same.

There is no difference between Protocol 1 and Protocol 2 at Bob's side. The
only difference between them is that in Protocol 1 Alice sends Bob each
coherent state one by one and in Protocol 2 Alice sends them all together.
The QM system that separates the big state into an $N$-state sequence can be
realized by storing the big state and sending each state one by one. In
Protocol 2, the QM system can be controlled by Eve. However, it can be seen
that if the QM system is put at Alice's side and cannot be controlled by
Eve, Protocol 2 is equivalent to Protocol 1. Therefore, the security of
Protocol 2 is weaker than that of Protocol 1. Therefore the secret key rate
of Protocol 1 is no less than that of Protocol 2.

\subsection{Protocol 3}

\begin{figure}[tbp]
\includegraphics[width=8cm]{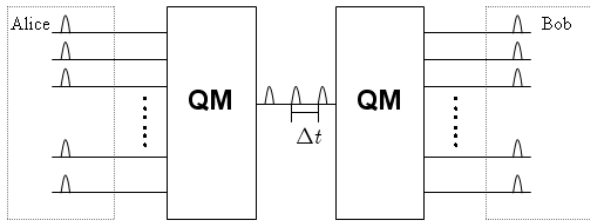}
\caption{Illustration of Protocol 3, where Bob's measurement equipment is
not shown in detail.}
\label{fig3}
\end{figure}

As shown in Fig. \ref{fig3}\, in step 2 of Protocol 2, Bob stores the
received pulses in another QM system. After receiving $N$ pluses, Bob reads
out the stored $N$ signals one by one through $N$ fibers and measures their
phase differences via M-Z systems which should be equivalent to the
measurements in Protocol 1 (these M-Z systems can be realized by the
detection equipments shown in Fig. \ref{fig4}).

It can be seen that if the QM system is at Bob's side and cannot be
controlled by Eve, then Protocol 3 is the same as Protocol 2. Since in
Protocol 3 this QM system can be controlled by Eve, the security of Protocol
3 is no stronger than that of Protocol 2.

\subsection{Protocol 4}

\begin{figure}[tbp]
\includegraphics[width=8cm]{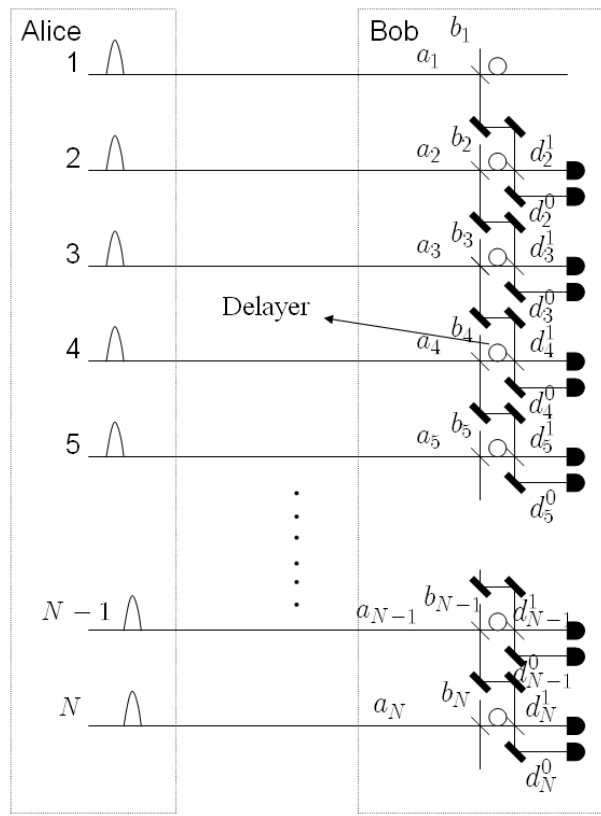}
\caption{Illustration of Protocol 4, where the delayer generates the time
delay that makes the two pulses meet at the beam splitter before the
detector. }
\label{fig4}
\end{figure}

In step 1, Alice generates a state $|\Psi _{\vec{x}}^{N}\rangle ={%
\bigotimes\limits_{i=kN+1}^{(k+1)N}}|(-1)^{x_{i}+1}\alpha \rangle $ and
sends it to Bob through $N$ fibers. In step 2, while receiving this big
state, Bob measures the phase difference between each two adjacent states at
the same time with the equipment shown in Fig. \ref{fig4}. Steps 3-9 remains
the same as in Protocol 1.

It can be seen that the only difference between Protocol 3 and Protocol 4 is
that in Protocol 3 Bob measures each phase difference one by one, but in
Protocol~4 Bob measures them all together. Since different detectors just
measure different field quadratures, the measurement operators that describe
these detectors commute with each other. Therefore, there is no difference
between measuring the phase difference one by one and measuring them all
together. Thus, Protocol 4 is equivalent to Protocol 3.

Since Protocol 4 is inferior to Protocol 1, in the following we will prove
the security of Protocol 4 first. Then the security of Protocol 1 follows.

\section{Security discussion}

Here we only discuss the security against the collective attack
under the infinite-key-length case. It means that Alice sends Bob
infinite number of the quantum state $|\Psi _{\vec{x}}^{N}\rangle $.
We remark that we do not place any restriction on the block length
$N$, which can be finite. Under the collective attack scenario Eve
attacks each state individually with the same superoperator and the
state Alice and Bob share is a state that can be written as a
product state. This means that after Alice sends Bob $M$ big states,
the state Alice, Bob and Eve share can be written as $(\rho
_{ABE}^{N})^{\otimes M}$, where $\rho _{ABE}^{N}$ is evolved from $|\Psi _{%
\vec{x}}^{N}\rangle $.\ Then it is enough for us to only discuss the
possible attacks to a single state $|\Psi _{\vec{x}}^{N}\rangle $ \cite%
{Renner2}. We also assume that quantum theory is correct, the device can be
trusted and Bob's detectors are ideal \cite{ideal}. In the following we
assume that $\vec{x}$, $\vec{y}$, $\vec{z}$, $\vec{l}$, $\vec{u}^{A}$ and $%
\vec{u}^{B}$ are the binary strings Alice and Bob obtain from one state $%
|\Psi _{\vec{x}}^{N}\rangle $, with $\vec{x}=(x_{1},x_{2},...,x_{N})$, $\vec{%
y}=(y_{1},y_{2},...y_{N})$, and similar for the others. Let $\vec{X}$, $\vec{%
Y} $, $\vec{Z}$, $\vec{L}$, $\vec{U}_{A}$ and $\vec{U}_{B}$ denote the
variables that can take the values $\vec{x}$, $\vec{y}$, $\vec{z}$, $\vec{l}$%
, $\vec{u}^{A}$ and $\vec{u}^{B}$, respectively. Let $X_{i}$, $Y_{i}$, $%
Z_{i} $, $L_{i}$, $U_{i}^{A}$, and $U_{i}^{B}$ denote the variables that can
take the values $x_{i}$, $y_{i}$, $z_{i}$, $l_{i}$, $u_{i}^{A}$, and $%
u_{i}^{B}$ respectively. In Ref. \cite{Renner,Deve}, it has been shown that
under the infinite-key-length case the secret key rate is given by the
difference between mutual informations. In the above protocols, $\vec{U}_{A}$
and $\vec{U}_{B}$ are the final variables that are used to distill the
secret keys. Then for Protocol 4, the secret key rate per big state is given
by
\begin{equation}
G\geq I(\vec{U}_{A}:\vec{U}_{B}|\vec{Z})-S(\vec{U}_{A}:E|\vec{Z})
\label{first key rate}
\end{equation}%
where $I$ and $S$ denote the Shannon mutual information \cite{Shannon} and
the Von Neuemann mutual information \cite{Nielson} respectively. Here the $I(%
\vec{U}_{A}:\vec{U}_{B}|\vec{Z})$ and $S(\vec{U}_{A}:E|\vec{Z})$ are
respectively given by%
\begin{equation}
\begin{array}{c}
I(\vec{U}_{A}:\vec{U}_{B}|\vec{Z})=\sum_{\vec{z}}P(\vec{z})I(\vec{U}_{A}:%
\vec{U}_{B}|\vec{Z}=\vec{z}) \\
S(\vec{U}_{A}:E|\vec{Z})=\sum_{\vec{z}}P(\vec{z})S(\vec{U}_{A}:E|\vec{Z}=%
\vec{z})%
\end{array}%
.  \label{mutual information}
\end{equation}

Before discussing the properties of these two mutual informations, we
introduce the following notations. We introduce vectors $w(\vec{z})$ and
$\vec{c}(\vec{z})$ 
in relation to Bob's counting results. Let $w(\vec{z})$ denote the weight of
$\vec{z}$, which gives the total number of $1$ in $\vec{z}$. Let $c_{i}(\vec{%
z})$ denotes the position of the $i$-th $1$ in the string $\vec{z}$.

From step 6, we see that $u_{i}^{A}=l_{i}\cdot z_{i}$. Since if $z_{i}=0$ we
have $u_{i}^{A}=0$, $\vec{u}^{A}$ can be given by $\vec{u}%
^{A}=(0,..,l_{c_{1}(\vec{z})},.0,..,l_{c_{w(\vec{z})}(\vec{z})},..,0)$. For
a given $\vec{z}$, we can change the order of elements and write $\vec{u}%
^{A} $ as $\vec{u}^{A}=(l_{c_{1}(\vec{z})},l_{c_{2}(\vec{z})},..,l_{c_{w(%
\vec{z})}(\vec{z})},0,..0)$. For convenience, we introduce $\vec{l}_{\vec{z}%
} $ and $\vec{y}_{\vec{z}}$, which are given by 
\begin{equation}
\begin{array}{c}
\vec{l}_{\vec{z}}=(l_{c_{1}(\vec{z})},l_{c_{2}(\vec{z})},..,l_{c_{w(\vec{z}%
)}(\vec{z})}) \\
\vec{y}_{\vec{z}}=(y_{c_{1}(\vec{z})},y_{c_{2}(\vec{z})},..,y_{c_{w(\vec{z}%
)}(\vec{z})})%
\end{array}%
.  \label{lz}
\end{equation}%
Then for a given $\vec{z}$, $\vec{u}^{A}$ can simply be given by
\begin{equation*}
\vec{u}^{A}=(\vec{l}_{\vec{z}},\vec{0})
\end{equation*}%
where $\vec{0}$ is the zero vector of length $N-w(\vec{z})$. Now we
introduce random variables $\vec{L}_{\vec{z}}$ and $\vec{Y}_{\vec{z}}$ of
length $w(\vec{z})$ 
that take on values $\vec{l}_{\vec{z}}$ and $\vec{y}_{\vec{z}}$:
\begin{equation}
\begin{array}{c}
\vec{L}_{\vec{z}}=(L_{c_{1}(\vec{z})},L_{c_{2}(\vec{z})},..,L_{c_{w(\vec{z}%
)}(\vec{z})}) \\
\vec{Y}_{\vec{z}}=(Y_{c_{1}(\vec{z})},Y_{c_{2}(\vec{z})},..,Y_{c_{w(\vec{z}%
)}(\vec{z})})%
\end{array}%
.  \label{LzYz}
\end{equation}%
Then for a given $\vec{z}$, $\vec{U}_{A}$ can be given by
\begin{equation}
\vec{U}_{A}=(\vec{L}_{\vec{z}},\vec{0})  \label{UA}
\end{equation}%
%
%
%
%
%
%
%
%
%
%
%
%
%
%
%
%
%
%
%
%
%
%
%
%
%
%
%
%
%
%
%
%
%
%
%
%
%
%
%
%
%
and similarly we have
\begin{equation}
\begin{array}{c}
\vec{u}^{B}=(\vec{y}_{\vec{z}},\vec{0}) \\
\vec{U}_{B}=(\vec{Y}_{\vec{z}},\vec{0})%
\end{array}%
.  \label{UB}
\end{equation}

Now we can rewrite the mutual information given by Eq. (\ref{mutual
information}) as follows.
\begin{gather}
I(\vec{U}_{A}:\vec{U}_{B}|\vec{Z})  \notag \\
=\sum_{\vec{z}}P(\vec{z})I(\vec{L}_{\vec{z}}\oplus \vec{0}:\vec{Y}_{\vec{z}%
}\oplus \vec{0}|\vec{Z}=\vec{z})  \notag \\
=\sum_{\vec{z}}P(\vec{z})[H(\vec{L}_{\vec{z}}\oplus \vec{0}|\vec{Z}=\vec{z})
\notag \\
+H(\vec{Y}_{\vec{z}}\oplus \vec{0}|\vec{Z}=\vec{z})-H(\vec{L}_{\vec{z}%
}\oplus \vec{0},\vec{Y}_{\vec{z}}\oplus \vec{0}|\vec{Z}=\vec{z})]  \notag \\
=\sum_{\vec{z}}P(\vec{z})I(\vec{L}_{\vec{z}}:\vec{Y}_{\vec{z}}|\vec{Z}=\vec{z%
})  \label{IAB}
\end{gather}%
where $H(\cdot )$ is Shannon entropy \cite{Shannon}, in the second line we
have applied the results given by Eq. (\ref{UA}) and (\ref{UB}), in the
third line and forth line we have used the definition of the Shannon mutual
information \cite{Shannon} and in the fifth line we have used the formula, $%
H(AB|C)=H(A|BC)+H(B|C)$, and the fact that the entropy of a given vector is
zero and $H(\cdot |\vec{Z}=\vec{z},\vec{0})=H(\cdot |\vec{Z}=\vec{z})$\ ($%
\vec{0}$\ is generated from $\vec{z}$). In the same way, we have
\begin{equation}
S(\vec{U}_{A}:E|\vec{Z})=\sum_{\vec{z}}P(\vec{z})S(\vec{L}_{\vec{z}}:E|\vec{Z%
}=\vec{z}).  \label{SAE}
\end{equation}%
Then from Eqs. (\ref{IAB}) and (\ref{SAE}), 
the final secret key rate per N pulses is given by
\begin{gather}
G\geq \sum_{\vec{z}}P(\vec{z})[I(\vec{L}_{\vec{z}}:\vec{Y}_{\vec{z}}|\vec{Z}=%
\vec{z})  \notag \\
-S(\vec{L}_{\vec{z}}:E|\vec{Z}=\vec{z})].  \label{another secret key rate}
\end{gather}

If Alice and Bob just discard all $l_{i}$\ and $y_{i}$ corresponding to $%
z_{i}=0$, and perform error correction and privacy amplification to the
sifted key, according to Ref. \cite{Renner} the secret key in this case is
also given by Eq. (\ref{another secret key rate}). Therefore, instead of
steps 6-9, Alice and Bob can simply introduce such post-selection step.

\section{Security proof}

Before giving the security proof of the DPS protocol, we will show the main
idea of our security proof. In the DPS protocol, what Alice sends to Bob are
two weak coherent states. Since these two states are non-orthogonal
Eve in principle cannot always distinguish them and cannot always know the
phase difference between two adjacent pulses. The mutual information per
pulse between Alice and Eve is thus less than one. Bob uses the M-Z system
to measure the phase difference between two adjacent pulses. This system can
let Bob definitely know the deterministic phase difference between the two
pulses with certain probability (Bob can definitely know the phase
difference if he gets a count.). Finally Alice and Bob only keep the
measurement results of the phase differences that Bob has good knowledge of.
The point is that either Bob knows that Eve has obtained good information on
a pulse pair (in which case he has not gotten a photon count and this result
is discarded) or Bob knows more about this result than Eve does (in which
case Alice and Bob proceed to distill a secret key by applying error
correction and privacy amplification to many of such result).
In the following, we will rigorously prove this when there is no bit error.
%
%
%
%
%
First, we prove that under the no-bit-error case Eve's state is uncorrelated
with the permutation of $\vec{z}$, which means that Eve's state is
independent with the position of Bob's detections. Then we show that Eve's
average information about the sifted keys is only determined by the total
number of Bob's detections. Since the maximum total information Eve gets from
Alice is restricted by Alice's modulation, through several properties of the
mutual information, we can give an upper bound to Eve's information on the
sifted data. Then the security of the DPS protocol under the no-bit-error
case is proved.

Here, we only limit our analysis to the no-bit-error case. For a collective
attack, it is enough to only discuss the attack for a single communication
\cite{Renner}. Therefore, in the following we only discuss the possible
attacks for a single state $|\Psi _{\vec{x}}^{N}\rangle $. Before giving the
security proof, we introduce the following notations. As shown in Fig. \ref%
{fig4}, the annihilation operator of the input fields on the $i$-th path at
Bob's side is denoted by $a_{i}$. The vacuum inject mode on the beam
splitter corresponding to the $i$-th fiber at Bob's side is denoted by $%
b_{i} $. The output modes to the detectors that detect the phase difference
between the $i$-th signal and $i-1$-th signal are described by $d_{i}^{0}$
and $d_{i}^{1}$, with 0 and 1 describing the lower and upper detectors
respectively (See Fig. \ref{fig4}). In the following, we denote the
detectors that correspond to the modes $d_{i}^{0}$ and $d_{i}^{1}$ by $%
DET_{i}^{0}$ and $DET_{i}^{1}$.

We assume that Bob and Eve's conditional states conditioned on Alice's
modulation is $\rho _{EB}^{N,\vec{x}}$. Now we will prove that if there is
no bit error, the reduced state $\rho _{E}^{N,\vec{x},\vec{z}}$, which
denotes Eve's conditional state after Bob's announcement of detected
signals, is independent of the permutation of $\vec{z}$. With this result
and several properties of the entropy, we can give a lower bound to Eq. (\ref%
{first key rate}).

\subsection{Permutation invariance of Eve's state to $\vec{z}$}

In this subsection we give the description of Bob's measurement first. Then
we prove a requirement implied by the no-bit-error condition. Finally we
prove that Eve's state is invariant under the permutation of $\vec{z}$.

From Fig. \ref{fig4}, we can see that the output modes $d_{i}^{0}$ and $%
d_{i}^{1}$ can respectively be given by%
\begin{equation}
\begin{array}{c}
d_{i}^{1}=\frac{1}{2}(a_{i}-a_{i-1})+\frac{i}{2}(b_{i}+b_{i-1}) \\
d_{i}^{0}=\frac{i}{2}(a_{i}+a_{i-1})+\frac{1}{2}(b_{i-1}-b_{i})%
\end{array}
.  \label{detection operater}
\end{equation}

To simplify Eq. (\ref{detection operater}) we introduce $\tilde{d}_{i}^{j}$
and 
$v_{i}^{j}$: 
\begin{equation}
\begin{array}{c}
\tilde{d}_{i}^{1}=\frac{1}{2}(a_{i}-a_{i-1}) \\
\tilde{d}_{i}^{0}=\frac{1}{2}(a_{i}+a_{i-1})%
\end{array}
\label{simplified}
\end{equation}%
%
%
%
%
%
%
%
%
%
%
%
%
%
%
%
%
%
%
%
%
%
%
%
%
%
%
%
%
%
%
%
%
%
%
%
%
%
%
%
%
%
%
%
%
and
\begin{equation}
\begin{array}{c}
vac_{i}^{1}=\frac{i}{2}(b_{i}+b_{i-1}) \\
vac_{i}^{0}=\frac{-i}{2}(b_{i-1}-b_{i})%
\end{array}%
.  \label{vacuum part}
\end{equation}

Since the phase factor $i$ can be absorbed into the annihilation operators,
the output modes $d_{i}^{0}$ and $d_{i}^{1}$ can simply be given by
\begin{equation}
\begin{array}{c}
d_{i}^{1}=\tilde{d}_{i}^{1}+vac_{j}^{1} \\
d_{i}^{0}=\tilde{d}_{i}^{0}+vac_{j}^{0}%
\end{array}%
.  \label{new expression}
\end{equation}

In our security analysis we can safely assume that Eve holds the
purification of $\rho _{EB}^{N,\vec{x}}$. This means that we can safely
assume that $\rho _{EB}^{N,\vec{x}}$ is a pure state \cite{Renner}. Let $%
|\Psi _{EB}^{N,\vec{x}}\rangle $ denote the pure conditional state Bob and
Eve share. According to the Schmidt decomposition (see, e.g., \cite{Nielson}%
), $|\Psi _{EB}^{N,\vec{x}}\rangle $ can be decomposed into several
orthogonal states,%
\begin{equation}
|\Psi _{EB}^{N,\vec{x}}\rangle ={\sum\limits_{k}}c_{k}^{\vec{x}}|\Phi
_{E,k}^{N,\vec{x}}\rangle |\Phi _{B,k}^{N,\vec{x}}\rangle .  \label{state}
\end{equation}

For convenience, in the following we denote the $\vec{l}$ generated from $%
\vec{x}$\ by $\vec{l}^{\vec{x}}$.

The no-bit-error condition can be described as follows. 
If there is no phase difference between the $i$-th and $i-1$-th states, in
principle the $DET_{i}^{1}$ detector should not generate a count; and if
there is phase difference, the $DET_{i}^{0}$ detector should not click. It
can be seen that this condition actually requires that the output mode $%
\tilde{d}_{i}^{\bar{l}_{i}^{\vec{x}}}$ of $\rho _{EB}^{N,\vec{x}}$
corresponds to the vacuum state, i.e., $d_{i}^{\bar{l}_{i}^{\vec{x}}}|\Psi
_{EB}^{N,\vec{x}}\rangle |vac\rangle =0$, where $\bar{l}_{i}^{\vec{x}}=l_{i}^{%
\vec{x}}\oplus 1$ and $|vac\rangle $ describes the vacuum state injected
through the $b_{i}$ mode. Since in Eq. (\ref{new expression}), the
annihilation operator $vac_{i}^{j}$ ($j=0,1$) acts on the vacuum state, we
can simply discard the vacuum mode. Then the no-bit-error condition can be
rewritten as
\begin{equation}
\tilde{d}_{i}^{\bar{l}_{i}^{\vec{x}}}|\Psi _{EB}^{N,\vec{x}}\rangle =0.
\label{zero property}
\end{equation}

From Eqs. (\ref{state}) and (\ref{simplified}), we know that Eq. (\ref{zero
property}) is equal to 
\begin{equation}
\lbrack a_{i}+(-1)^{l_{i}^{\vec{x}}+1}a_{i-1}]|\Phi _{B,k}^{N,\vec{x}%
}\rangle =0,  \label{equality}
\end{equation}%
for arbitrary $i>1$ and $k$. Eq. (\ref{equality}) gives the no-bit-error
condition. 
This analysis can be summarized by the following Lemma.

\textit{Lemma 1}: If there is no bit error between Alice and Bob, Bob's
state satisfies

\begin{equation}
a_{i}|\Phi _{B,k}^{N,\vec{x}}\rangle =(-1)^{l_{i}^{\vec{x}}}a_{i-1}|\Phi
_{B,k}^{N,\vec{x}}\rangle .  \label{equality2}
\end{equation}

Since $\tilde{d}_{i}^{l_{i}}=\frac{1}{2}[a_{i}+(-1)^{l_{i}}a_{i-1}]$, from
Eq. (\ref{equality2}), we get%
\begin{equation}
\tilde{d}_{i}^{l_{i}^{\vec{x}}}|\Phi _{B,k}^{N,\vec{x}}\rangle =a_{i}|\Phi
_{B,k}^{N,\vec{x}}\rangle .  \label{equality3}
\end{equation}%
Eqs. (\ref{zero property}) and (\ref{equality3}) give us the relationship
between the input modes and the output modes of the M-Z system under the
no-bit-error case. With this Lemma we can prove the permutation invariance
of Eve's state to $\vec{z}$ finally.

Combining Eqs. (\ref{equality2}) and (\ref{equality3}) and the result that $%
l_{i}^{\vec{x}}=x_{i}\oplus x_{i-1}$, we have
\begin{equation}
(-1)^{x_{i}}\tilde{d}_{i}^{l_{i}^{\vec{x}}}|\Phi _{B,k}^{N,\vec{x}}\rangle
=(-1)^{x_{i^{\prime }}}\tilde{d}_{i^{\prime }}^{l_{i^{\prime }}^{\vec{x}%
}}|\Phi _{B,k}^{N,\vec{x}}\rangle  \label{detector}
\end{equation}%
%
%
%
%
%
%
%
%
%
%
%
%
%
%
%
%
%
%
%
%
%
%
%
%
%
%
%
%
%
%
%
%
%
%
%
%
%
%
%
%
%
%
%
%
for arbitrary $i$, $i^{\prime }$ and $k$. It can be seen that
Eq. (\ref{detector}) can be generalized to
\begin{eqnarray}
&&(-1)^{x_{i_{1}+}x_{i_{2}}+...+x_{i_{q}}}\tilde{d}_{i_{1}}^{l_{i_{1}}^{\vec{%
x}}}\tilde{d}_{i_{2}}^{l_{i_{2}}^{\vec{x}}}...\tilde{d}_{i_{q}}^{l_{i_{q}}^{%
\vec{x}}}|\Phi _{B,k}^{N,\vec{x}}\rangle  \label{key} \\
&=&(-1)^{x_{i_{1}^{\prime }+}x_{i_{2}^{\prime }}+...+x_{i_{q}^{\prime }}}%
\tilde{d}_{i_{1}^{\prime }}^{l_{i_{1}^{\prime }}^{\vec{x}}}\tilde{d}%
_{i_{2}^{\prime }}^{l_{i_{2}^{\prime }}^{\vec{x}}}...\tilde{d}%
_{i_{q}^{\prime }}^{l_{i_{q}^{\prime }}^{\vec{x}}}|\Phi _{B,k}^{N,\vec{x}%
}\rangle  \notag
\end{eqnarray}%
for arbitrary integer $i_{p}>1$, $i_{p}^{\prime }>1$, $k$ and $q$, where the
subscription $p=1,\ldots ,q$ ($p$ comes from $i_{p}$).

Bob's measurement is a projection. The detector $DET_{i}^{j}$ maps the
received state into the photon number space $\{|vac\rangle \langle
vac|,|1\rangle \langle 1|,...\}$. This projection subspace of $DET_{i}^{j}$
can be given by
\begin{equation}
\text{Map}_{i}^{j}=\{M_{iv}^{j}=\frac{1}{v!}(d_{i}^{j+})^{v}|vac\rangle
\langle vac|(d_{i}^{j})^{v},|v=0,1,...\},  \label{M}
\end{equation}%
where $v$ denotes the number of photons received by the detector $%
DET_{i}^{j} $.

Since $d_{i}^{j}$ is composed by an operator acting on the received state
and an operator acting on the vacuum state, from Eqs. (\ref{new expression}%
), (\ref{key}) and (\ref{M}), it can be seen that for arbitrary integer $i$,
$i^{\prime }$, $v$ and $k$ we always have
\begin{equation}
\langle vac|\langle \Phi _{B,k}^{N,\vec{x}}|M_{iv}^{l_{i}^{\vec{x}}}|\Phi
_{B,k}^{N,\vec{x}}\rangle |vac\rangle =\langle vac|\langle \Phi _{B,k}^{N,\vec{x}%
}|M_{i^{\prime }v}^{l_{i^{\prime }}^{\vec{x}}}|\Phi _{B,k}^{N,\vec{x}%
}\rangle |vac\rangle  \label{M relation}
\end{equation}%
where $|vac\rangle $ denotes the vacuum state injected through the $b_{i}^{j}$
mode and this equality holds for $v=0$ because the modulus of the projection
to the vacuum state plus that to other states equals to one. Since $%
|vac\rangle $ is always a vacuum state, in the following we just leave it out
in our expression for convenience. Eq. (\ref{M relation}) can be expended to
the multi-detection case as follows: 
\begin{eqnarray}
&&\langle \Phi _{B,k}^{N,\vec{x}}|M_{i_{1}v_{1}}^{l_{i_{1}}^{\vec{x}}}\cdots
M_{i_{q}v_{q}}^{l_{i_{q}}^{\vec{x}}}|\Phi _{B,k}^{N,\vec{x}}\rangle
\label{Multi M} \\
&=&\langle \Phi _{B,k}^{N,\vec{x}}|M_{i_{1}^{\prime
}v_{1}}^{l_{i_{1}^{\prime }}^{\vec{x}}}\cdots M_{i_{q}^{\prime
}v_{q}}^{l_{i_{q}^{\prime }}^{\vec{x}}}|\Phi _{B,k}^{N,\vec{x}}\rangle
\notag
\end{eqnarray}%
where we have introduced subscripts $1,2,...,$ and $q$ to enumerate
different operators and have omitted the vacuum state part.

From Eqs. (\ref{state}) and (\ref{Multi M}) it can be obtained that
\begin{eqnarray}
&&\langle \Psi _{EB}^{N,\vec{x}}|M_{i_{1}v_{1}}^{l_{i_{1}}^{\vec{x}}}\cdots
M_{i_{q}v_{q}}^{l_{i_{q}}^{\vec{x}}}|\Psi _{EB}^{N,\vec{x}}\rangle  \notag \\
&=&\langle \Psi _{EB}^{N,\vec{x}}|M_{i_{1}^{\prime }v_{1}}^{l_{i_{1}^{\prime
}}^{\vec{x}}}\cdots M_{i_{q}^{\prime }v_{q}}^{l_{i_{q}^{\prime }}^{\vec{x}%
}}|\Psi _{EB}^{N,\vec{x}}\rangle .  \label{probability}
\end{eqnarray}%
Also from Eqs. (\ref{zero property}) and (\ref{M}) it can be seen that if $%
j\neq l_{i}^{\vec{x}}$ and $v>0$ we always have%
\begin{equation}
\langle \Phi _{B,k}^{N,\vec{x}}|M_{iv}^{j}|\Phi _{B,k}^{N,\vec{x}}\rangle =0.
\label{zero}
\end{equation}

With the above results, we can evaluate Eve's conditional state conditioned
on Bob's announcement. We know that if Bob maps his state into the subspace $%
M=|\beta \rangle \langle \beta |$ (for convenience we denote it by $\beta $)
then Eve's state will collapse to the state
\begin{equation}
\rho _{E}^{N,\vec{x},\beta }=\frac{tr_{B}M\rho _{EB}^{N,\vec{x}}M^{+}}{%
tr_{EB}M\rho _{EB}^{N,\vec{x}}M^{+}},  \label{reduced state}
\end{equation}%
where $\rho _{EB}^{N,\vec{x}}$ is the state Eve and Bob share before Bob's
measurement \cite{Nielson}. In our security analysis we can consider the
worst case (favoring Eve), in which the state $\rho _{EB}^{N,\vec{x}}$ is
pure and described by $|\Psi _{EB}^{N,\vec{x}}\rangle $. For the pure state
case, Eq. (\ref{reduced state}) becomes
\begin{equation}
|\Psi _{E}^{N,\vec{x},\beta }\rangle ={\sum\limits_{k}}\frac{c_{k}^{\vec{x}}%
\sqrt{\langle \Phi _{B,k}^{N,\vec{x}}|M|\Phi _{B,k}^{N,\vec{x}}\rangle }%
|\Phi _{E,k}^{N,\vec{x}}\rangle }{\sqrt{\langle \Psi _{EB}^{N,\vec{x}%
}|M|\Psi _{EB}^{N,\vec{x}}\rangle }}  \label{pure reduced}
\end{equation}%
where we have used Eq. (\ref{state}) and the results that $M=M^{+}=MM^{+}$.

Bob's announcement $\vec{z}$\ corresponds to several possible orthogonal
subspaces given by
\begin{eqnarray}
\{S_{\vec{z},\vec{v}>0}^{\vec{j}} &=&M_{c_{1}(\vec{z}%
),v_{1}}^{j_{1}}M_{c_{2}(\vec{z}),v_{2}}^{j_{2}}\cdots M_{c_{w(\vec{z})}(%
\vec{z}),v_{w(\vec{z})}}^{j_{w(\vec{z})}}
\notag
\\
&&\otimes |vac\rangle \langle vac|
\big\vert j_{q} =0,1;v_{q}>0\},  \label{S expression}
\end{eqnarray}%
where 
$\vec{j}=\{j_{1},j_{2},...,j_{w(\vec{z})}\}$, $\vec{v}%
=\{v_{1},v_{2},...,v_{w(\vec{z})}\}$, $\vec{v}>0$ means all of its element $%
v_{q}$ is larger than zero, the element $M_{c_{q}(\vec{z}),v_{q}}^{j_{q}}$
corresponds to the measurement result that detector $DET_{c_{q}(\vec{z}%
)}^{j_{q}}$ receives $v_{q}$ photons and $|vac\rangle \langle vac|$ denotes
other detectors have not received any photon. The set of all operators $S_{%
\vec{z},\vec{v}>0}^{\vec{j}}$'s only spans a subspace. From Fig. \ref{fig4}
we can see that one mode coming from the first pulse and one mode coming
from the $N$-th pulse are not detected by Bob (or the measurement results of
them are discarded by Bob). Therefore we need to also consider Bob's
undetected subspace. We assume Bob's undetected subspace (or the measurement
results of which are discarded) is spanned by the orthogonal basis \{$%
|R_{1}\rangle ,|R_{2}\rangle ,...$\}. Then all $S_{\vec{z},\vec{v}>0}^{\vec{j%
}}$'s and $|R_{i}\rangle $'s make up a complete set of projections for Bob.
An announcement $\vec{z}$\ corresponds to several measurement results. Each
result corresponds to a collapsed state of Eve. Then an announcement $\vec{z}
$ collapses Eve's state into a mixed state made up by several pure state.
For convenience, we let $R_{i}$ denote the projector $|R_{i}\rangle \langle
R_{i}|$. Then $S_{\vec{z},\vec{v}>0}^{\vec{j}}$ is orthogonal to $R_{i}$.
From Eqs. (\ref{lz}) and (\ref{zero}) we know that for a given $\vec{x}$, $%
\vec{z}$, if $\vec{j}\neq \vec{l}_{\vec{z}}^{\vec{x}}$, it is always
satisfied that $\langle \Phi _{B,k}^{N,\vec{x}}|S_{\vec{z},\vec{v}>0}^{\vec{j%
}}|\Phi _{B,k}^{N,\vec{x}}\rangle =0$ and $\langle \Phi _{B,k}^{N,\vec{x}%
}|S_{\vec{z},\vec{v}>0}^{\vec{j}}R_{i}|\Phi _{B,k}^{N,\vec{x}}\rangle =0$,
where the second equality comes from the fact that $S_{\vec{z},\vec{v}>0}^{%
\vec{j}}$ is orthogonal to $R_{i}$. Then after we generalized Eq. (\ref{pure
reduced}) to the multi-operator and multi-measurement-result case, we know
that if Bob's announcement is $\vec{z}$, then Eve's conditional state becomes%
\begin{eqnarray}
\rho _{E}^{N,\vec{x},\vec{z}} &=&{\sum\limits_{\vec{v}>0,i}}\frac{\langle
\Psi _{EB}^{N,\vec{x}}|S_{\vec{z},\vec{v}>0}^{\vec{l}_{\vec{z}}^{\vec{x}%
}}R_{i}|\Psi _{EB}^{N,\vec{x}}\rangle }{{P(\vec{z}|\vec{x})}}  \label{roh} \\
&&\cdot \text{Proj}({\sum\limits_{k}}\frac{c_{k}^{\vec{x}}\sqrt{\langle \Phi
_{B,k}^{N,\vec{x}}|S_{\vec{z},\vec{v}>0}^{\vec{l}_{\vec{z}}^{\vec{x}%
}}R_{i}|\Phi _{B,k}^{N,\vec{x}}\rangle }|\Phi _{E,k}^{N,\vec{x}}\rangle }{%
\sqrt{\langle \Psi _{EB}^{N,\vec{x}}|S_{\vec{z},\vec{v}>0}^{\vec{l}_{\vec{z}%
}^{\vec{x}}}R_{i}|\Psi _{EB}^{N,\vec{x}}\rangle }})  \notag
\end{eqnarray}%
where $\text{Proj}(|\Psi \rangle )$ denotes the state $|\Psi \rangle \langle
\Psi |$,
\begin{equation}
{P(\vec{z}|\vec{x})=\sum\limits_{\vec{v}>0}}\langle \Psi _{EB}^{N,\vec{x}%
}|S_{\vec{z},\vec{v}>0}^{\vec{l}_{\vec{z}}^{\vec{x}}}|\Psi _{EB}^{N,\vec{x}%
}\rangle  \label{Pzx}
\end{equation}%
describes the conditional probability of announcement ${\vec{z}}$\ for a
given ${\vec{x}}$, $\langle \Psi _{EB}^{N,\vec{x}}|S_{\vec{z},\vec{v}>0}^{%
\vec{l}_{\vec{z}}^{\vec{x}}}|\Psi _{EB}^{N,\vec{x}}\rangle $\ gives the
probability of the measurement result corresponding to the operator $S_{\vec{%
z},\vec{v}>0}^{\vec{l}_{\vec{z}}^{\vec{x}}}$ and we have used the result
that if $\vec{j}\neq \vec{l}_{\vec{z}}^{\vec{x}}$, then
$\langle \Psi _{EB}^{N,\vec{x}}|S_{\vec{z},\vec{v}>0}^{\vec{j}}R_{i}|\Psi
_{EB}^{N,\vec{x}}\rangle =0$.\

It can be seen that for the detector $DET_{i}^{j}$ the projector to the
vacuum state, $|vac\rangle \langle vac|$, can also be written as $M_{i0}^{j}$.
Then from Eqs. (\ref{Multi M}) and (\ref{S expression}), it can be seen that
if $w({\vec{z}})=w({\vec{z}}^{\prime })$, then 
\begin{equation}
\begin{array}{c}
\langle \Phi _{B,k}^{N,\vec{x}}|S_{\vec{z},\vec{v}>0}^{\vec{l}_{\vec{z}}^{%
\vec{x}}}|\Phi _{B,k}^{N,\vec{x}}\rangle =\langle \Phi _{B,k}^{N,\vec{x}}|S_{%
\vec{z}^{\prime },\vec{v}>0}^{\vec{l}_{\vec{z}^{\prime }}^{\vec{x}}}|\Phi
_{B,k}^{N,\vec{x}}\rangle , \\
\langle \Phi _{B,k}^{N,\vec{x}}|S_{\vec{z},\vec{v}>0}^{\vec{l}_{\vec{z}}^{%
\vec{x}}}R_{i}|\Phi _{B,k}^{N,\vec{x}}\rangle =\langle \Phi _{B,k}^{N,\vec{x}%
}|S_{\vec{z}^{\prime },\vec{v}>0}^{\vec{l}_{\vec{z}^{\prime }}^{\vec{x}%
}}R_{i}|\Phi _{B,k}^{N,\vec{x}}\rangle ,%
\end{array}
\label{s relation}
\end{equation}%
where we have used results that the output mode $d_{j}^{\bar{l}_{i}^{\vec{x}%
}}$ is a vacuum state and $S_{\vec{z},\vec{v}>0}^{\vec{j}}$ is orthogonal to
$R_{i}$.

We know that ${P(\vec{z},\vec{x})=P(\vec{z}|\vec{x})\cdot P(\vec{x})}$, $P(%
\vec{x}|\vec{z})=P(\vec{z},\vec{x})/P(\vec{z})$ and $P(\vec{z})=\sum_{\vec{x}%
}{P(\vec{z},\vec{x})}$, so from Eqs. (\ref{state}) (\ref{Pzx}) and (\ref{s
relation}) it can be obtained that
\begin{equation}
\begin{array}{c}
P(\vec{z}|\vec{x})=P(\vec{z}^{\prime }|\vec{x}) \\
{P(\vec{z},\vec{x})=P(\vec{z}}^{\prime },{\vec{x})} \\
P(\vec{z})=P(\vec{z}^{\prime }) \\
P(\vec{x}|\vec{z})=P(\vec{x}|\vec{z}^{\prime })%
\end{array}
\label{pz equality}
\end{equation}%
for $w({\vec{z}})=w({\vec{z}}^{\prime })$.

By combining Eqs. (\ref{probability}), (\ref{roh}), (\ref{Pzx}), (\ref{s
relation}) and (\ref{pz equality}) we obtain our final result:

\textit{Lemma 2}: If there is no bit error and $w({\vec{z}})=w({\vec{z}}%
^{\prime })$, we always have
\begin{equation}
\begin{array}{c}
\rho _{E}^{N,\vec{x},\vec{z}}=\rho _{E}^{N,\vec{x},\vec{z}^{\prime }} \\
\rho _{E}^{N,\vec{z}}=\rho _{E}^{N,\vec{z}^{\prime }}%
\end{array}%
.  \label{symmetry}
\end{equation}

The second equation in the Eq. (\ref{symmetry}) is derived from Eq. (\ref{pz
equality}) and the fact that $\rho _{E}^{N,\vec{z}}=\sum_{\vec{x}}P(\vec{x}|%
\vec{z})\rho _{E}^{N,\vec{x},\vec{z}}$. Eq. (\ref{symmetry}) shows that
Eve's conditional state is only determined by the total number of Bob's
detections and is independent of the positions of the detections. This means
that 
only the weight of ${\vec{z}}$\ matters, and we can use this instead of ${%
\vec{z}}$. In the following we introduce $\rho _{E}^{N,\vec{x},t}$
to denote the same conditional state that Eve holds for any ${\vec{z}}$ with
weight $t$, i.e.,
\begin{equation}
\begin{array}{c}
t=w(\vec{z}) \\
\rho _{E}^{N,\vec{x},t}=\rho _{E}^{N,\vec{x},\vec{z}}%
\end{array}%
.  \label{t roh}
\end{equation}%
%
%
%
%
%
%
%
%
%
%
%
%
%
%
%
%
%
%
%
%
%
%
%
%
%
%
%
%
%
%
%
%
%
%
%
%
%
%
%
%
%
%
%
%
Here $\rho _{E}^{N,\vec{x},t}$\ can be regarded as Eve's conditional state
while Bob only announces the total number of his detections.

With this new notation, we can simplify the expression of the mutual
information discussed in the following.

\subsection{Discussion of Eve's information}

Since Eve's state is invariant under the permutation of $\vec{z}$, it is
possible that Eve's information about Alice's bit string can also be bounded
by a term that only depends on Bob's total detection count (the weight of $%
\vec{z}$). In this subsection we upper bound Eve's information about Alice's
bit string by a term that is only conditioned on Bob's total detection count
by applying the super-subadditivity result proved in Appendix A. Then we
give restrictions on Eve's information by using the fact that the maximum
information Eve can get on Alice's bit string should be no larger than the
Holevo quantity \cite{Nielson} of the states sent by Alice and the fact that
the conditional mutual information should be no larger than the entropy of
Alice's modulation. With these restrictions it is possible for us to give an
upper bound to Eve's information about Alice's bit string.

First we define%
\begin{equation}
F(w):={\sum\limits_{\vec{z}}}P(\vec{z}|W=w)S(\vec{L}_{\vec{z}}:E|\vec{Z}=%
\vec{z})  \label{Def Fw}
\end{equation}%
where we have introduced a variable $W$ that can take the value of $w(\vec{z}%
)$ and $P(\vec{z}|W=w)$ denotes the probability of $\vec{z}$ when $w(\vec{z}%
)=w$.

From Eq. (\ref{SAE}) we know that Eve's information about Alice's bit string
can be rewritten as

\begin{equation}
S(\vec{U}_{A}:E|\vec{Z})=\sum_{w=1}^{N-1}P(w)F(w) .  \label{ReDef SAE}
\end{equation}

Now we will see that the $F(w)$ is no larger than a term that only depends
on the total count $w$.

According to the definition of Von Neumann mutual information \cite{Nielson}%
, the mutual information between Alice and Eve conditioned on Bob's
announcement $\vec{z}$\ can be given by

\begin{equation}
S(\vec{L}_{\vec{z}}:E|\vec{Z}=\vec{z})=S(\rho _{E}^{N,\vec{z}})-{%
\sum\limits_{\vec{l}_{\vec{z}}}}P(\vec{l}_{\vec{z}}|\vec{z})S(\rho _{E}^{N,%
\vec{l}_{\vec{z}},\vec{z}})  \label{SEAZ}
\end{equation}%
where $P(\vec{l}_{\vec{z}}|\vec{z})$ is the probability that $\vec{L}_{\vec{z%
}}$ takes the value of $\vec{l}_{\vec{z}}$ while $\vec{Z}=\vec{z}$,
\begin{equation*}
\rho _{E}^{N,\vec{l}_{\vec{z}},\vec{z}}={\sum\limits_{\vec{x}\in \Gamma (%
\vec{l}_{\vec{z}})}P(\vec{x}|\vec{l}_{\vec{z}},\vec{z})}\rho _{E}^{N,\vec{x},%
\vec{z}},
\end{equation*}%
$\Gamma (\vec{l}_{\vec{z}})$ is the collection of all $\vec{x}$'s that
satisfy $x_{c_{i}(\vec{z})}\oplus x_{c_{i}(\vec{z})-1}=(l_{\vec{z}})_{i}$
and ${P(\vec{x}|\vec{l}_{\vec{z}},\vec{z})}$ is the probability that $\vec{X}
$ takes the value of ${\vec{x}}$ while $\vec{L}_{\vec{z}}=\vec{l}_{\vec{z}}$
and $\vec{Z}=\vec{z}$.

Also From Eq. (\ref{pz equality}) we know that the conditional probability $%
P(\vec{x}|\vec{z})$ is actually only conditioned on the weight of $\vec{z}$,
so we can introduce $P(\vec{x}|w(\vec{z}))$ to denote $P(\vec{x}|\vec{z})$.
Since $\vec{l}_{\vec{z}}$ is part of $\vec{l}$ and $\vec{l}$ is generated
from $\vec{x}$, the conditional probability $P(\vec{l}_{\vec{z}}|\vec{z})$
is also only conditioned on the weight of $\vec{z}$. Therefore we can also
write $P(\vec{l}_{\vec{z}}|\vec{z})$ as $P(\vec{l}_{\vec{z}}|w(\vec{z}))$
and ${P(\vec{x}|\vec{l}_{\vec{z}},\vec{z})}$ as ${P(\vec{x}|\vec{l}_{\vec{z}%
},w(\vec{z}))}$, where ${P(\vec{x}|\vec{l}_{\vec{z}},\vec{z})=P(\vec{x},\vec{%
l}_{\vec{z}}|\vec{z})/P(\vec{l}_{\vec{z}}|\vec{z})}$, and $P(\vec{l}_{\vec{z}%
}|w(\vec{z}))$\ denotes the probability that $\vec{L}_{\vec{z}}$\ takes the
value of $\vec{l}_{\vec{z}}$\ while $W=w(\vec{z})$. Then from Lemma 2 we
know that $\rho _{E}^{N,\vec{l}_{\vec{z}},\vec{z}}$ can be written as $\rho
_{E}^{N,\vec{l}_{\vec{z}},w(\vec{z})}$. Here, that the conditional
probabilities, $P(\vec{l}_{\vec{z}}|w(\vec{z}))$ and $P(\vec{x}|\vec{l}_{%
\vec{z}},w(\vec{z}))$, and the conditional state $\rho _{E}^{N,\vec{l}_{\vec{%
z}},w(\vec{z})}$\ are conditioned on the weight of $\vec{z}$ means that they
are conditioned on 
Bob's announcement of the total number of his detection count rather than
the actual detection positions.

Now we can use Lemma 2 to simplify $S(\vec{L}_{\vec{z}}:E|\vec{Z}=\vec{z})$\
and thus $F(w)$. From Lemma 2 and Eq. (\ref{t roh}) we know that $\rho
_{E}^{N,\vec{l}_{\vec{z}},\vec{z}}$ and $\rho _{E}^{N,\vec{z}}$ in Eq. (\ref%
{SEAZ}) can be replaced by $\rho _{E}^{N,\vec{l}_{\vec{z}},w(\vec{z})}$ and $%
\rho _{E}^{N,w(\vec{z})}$. Then $S(\vec{L}_{\vec{z}}:E|\vec{Z}=\vec{z})$
becomes%
\begin{eqnarray}
S(\vec{L}_{\vec{z}} &:&E|\vec{Z}=\vec{z})=S(\rho _{E}^{N,w(\vec{z})})-{%
\sum\limits_{\vec{l}_{\vec{z}}}}P(\vec{l}_{\vec{z}}|\vec{z})S(\rho _{E}^{N,%
\vec{l}_{\vec{z}},w(\vec{z})})  \notag \\
&=&S(\vec{L}_{\vec{z}}:E|W=w(\vec{z}))  \label{SLET}
\end{eqnarray}%
where we have used the definition of the conditional Von Neumann mutual
information which says
\begin{equation}
S(\vec{L}_{\vec{z}}:E|W=w)=S(\rho _{E}^{N,w})-{\sum\limits_{\vec{l}_{\vec{z}%
}}}P(\vec{l}_{\vec{z}}|w)S(\rho _{E}^{N,\vec{l}_{\vec{z}},w})
\label{SLET def}
\end{equation}%
and the result $P(\vec{l}_{\vec{z}}|w)=P(\vec{l}_{\vec{z}}|\vec{z})$ for $%
w=w(\vec{z})$. Here $S(\vec{L}_{\vec{z}}:E|W=w)$\ is actually Eve's
information about $\vec{L}_{\vec{z}}$ while Bob only announces the total
number of his counts. It can be seen that Eve's information about certain $%
\vec{L}_{\vec{z}}$\ does not change if Bob publishes detailed position of
his detection or only the total number of his detection.

We know that the mutual information $S(\vec{L}_{\vec{z}}:E|W=w(\vec{z}))$
can also be given by%
\begin{gather}
S(\vec{L}_{\vec{z}}:E|W=w(\vec{z}))  \label{re-expression} \\
=H(\vec{L}_{\vec{z}}|W=w(\vec{z}))-S(\vec{L}_{\vec{z}}|E,W=w(\vec{z}))
\notag \\
\leq w(\vec{z})-S(\vec{L}_{\vec{z}}|E,W=w(\vec{z}))  \notag
\end{gather}%
where we have used the fact that $H(\vec{L}_{\vec{z}}|W=w(\vec{z}))\leq w(%
\vec{z})$.

Then by inserting Eqs. (\ref{SLET}) and (\ref{re-expression}) to the
expression of $F(w)$ given by Eq. (\ref{Def Fw}) we can immediately get%
\begin{equation}
F(w)\leq w-{\sum\limits_{\vec{z}}}P(\vec{z}|W=w)S(\vec{L}_{\vec{z}}|E,W=w)
\label{summation}
\end{equation}%
where we have used the fact that ${\sum\limits_{\vec{z}}}P(\vec{z}|W=w)=1$.
In the following it can be seen that the last term in Eq. (\ref{summation})
can be lower bounded by a term depending on Bob's total count, $w$.

The probability $P(\vec{z}|W=w)$ for $w(\vec{z})=w$\ in Eq. (\ref{summation}%
) can be given by
\begin{eqnarray}
P(\vec{z}|W &=&w)=\frac{P(\vec{z})}{\sum\limits_{\vec{z}^{\prime }|w(\vec{z}%
^{\prime })=w}P(\vec{z}^{\prime })}  \notag \\
&=&\frac{1}{\sum\limits_{\vec{z}^{\prime }|w(\vec{z}^{\prime })=w}1}=\frac{1%
}{C_{N-1}^{w}}  \label{Pzw}
\end{eqnarray}%
where in the second line we have used the Eq. (\ref{pz equality}), and $%
C_{N-1}^{w}=\sum\limits_{\vec{z}^{\prime }|w(\vec{z}^{\prime })=w}1$ denotes
the number of permutations of $w$ in $N-1$.

Since $\vec{L}_{\vec{z}}$ can be regarded as $w(\vec{z})$ selection from $%
\vec{L}$, which is composed of $N-1$ elements, by using 
super-subadditivity of entropy (which is proved in Appendix A), we have
\begin{equation}
{\sum\limits_{\vec{z}|w(\vec{z})=w}}S(\vec{L}_{\vec{z}}|E,W=w)\geq \frac{%
wC_{N-1}^{w}}{N-1}S(\vec{L}|E,W=w).  \label{subadditivity}
\end{equation}%
Then by combining Eqs. (\ref{summation}), (\ref{Pzw}) and (\ref%
{subadditivity}), we can obtain
\begin{equation}
{F}(w)\leq w-\frac{w}{N-1}S(\vec{L}|E,W=w).  \label{Fw}
\end{equation}

Finally, from Eqs. (\ref{ReDef SAE}) and (\ref{Fw}) we know that Eve's
information about Alice's bit string satisfies%
\begin{gather}
S(\vec{U}_{A}:E|\vec{Z})=\sum_{w=1}^{N-1}P(w)F_{w}  \notag \\
\leq \sum_{w=1}^{N-1}P(w)w-\frac{1}{N-1}\sum_{w=1}^{N-1}P(w)wS(\vec{L}|E,W=w)
\label{Eves information}
\end{gather}%
where we sum over $w$ in the range of 1 to $N-1$, because the maximum value
of $w$ is $N-1$ and Alice and Bob discard the result while there is no
detection.

Eq. (\ref{Eves information}) gives us an upper bound on Eve's information
about Alice's state, when there is no bit error. This upper bound no longer
depends on Bob's counting positions denoted by $\vec{z}$. Now the remaining
problem is to find some practical restrictions, so that the upper bound of
Eve's information can be calculated.

Since all possible $w$\ is within $0$ and $N-1$, it can be seen that
\begin{equation}
\sum_{w=0}^{N-1}P(w)S(\vec{L}|E,W=w)=S(\vec{L}|E,W) .  \label{upper bound}
\end{equation}%
In the following we will see that if $S(\vec{L}|E,W)$ can be lower bounded,
then $S(\vec{U}_{A}:E|\vec{Z})$ can be upper bounded.

According to the definition of mutual information, $S(\vec{L}|E,W)$ can be
expressed as
\begin{eqnarray}
S(\vec{L}|E,W) &=&H(\vec{L})-S(\vec{L}:E,W)  \notag \\
&=&N-1-S(\vec{L}:E,W)  \label{conditional entropy}
\end{eqnarray}%
where $H(\vec{L})=N-1$, and $S(\vec{L}:E,W)$ is the mutual information
between $\vec{L}$ and the combination of Eve and system $W$.

Since $\vec{L}$ is generated from $X$ and $l_{i}=x_{c_{i}}\oplus x_{c_{i}-1}$%
, given $x_{1}$ and $\vec{l}$, we can completely reconstruct $\vec{x}$.
Therefore, we have 
\begin{equation}
S(\vec{L},X_{1}:E,W)=S(\vec{X}:E,W).  \label{SLXEW}
\end{equation}%
Because discarding a subsystem never increases the mutual information \cite%
{Nielson}, Eq. (\ref{SLXEW}) leads to
\begin{equation}
S(\vec{L}:E,W)\leq S(\vec{X}:E,W).  \label{SLEW}
\end{equation}%
The term $S(\vec{X}:E,W)$ is the mutual information between Alice and the
combination of Eve and system $W$. The system $E$, $W$ can be regarded as a
system locally generated from the original state, $\rho _{A}^{\otimes
N}=\sum_{\vec{x}}P(\vec{x})|\Psi _{\vec{x}}^{N}\rangle \langle \Psi _{\vec{x}%
}^{N}|$, sent out by Alice. Since local operations cannot increase the
mutual information, the maximum mutual information $S(\vec{X}:E,W)$ should
not be bigger than the maximum information one can obtain from the original
state $\rho _{A}^{\otimes N}$, which is upper bounded by the Holevo quantity
of $\rho _{A}^{\otimes N}$ \cite{Nielson}. The Holevo quantity of $\rho
_{A}^{\otimes N}\ $\ is given by $NS(A)$, where
\begin{equation}
S(A)=h[\frac{1}{2}(1-|\langle -\alpha |\alpha \rangle |)]  \label{SA}
\end{equation}%
is the entropy of a single state $\rho _{A}=\frac{1}{2}|-\alpha \rangle
\langle -\alpha |+\frac{1}{2}|\alpha \rangle \langle \alpha |$ and $%
h(x)=-x\log _{2}x-(1-x)\log _{2}(1-x)$ is the binary Shannon entropy
function \cite{Shannon,Nielson}. Now we know that
\begin{equation}
S(\vec{L}:E,W)\leq NS(A).  \label{NSA}
\end{equation}%
Then from Eqs. (\ref{upper bound}), (\ref{conditional entropy}) and (\ref%
{NSA}), we can derive the final constraint on
Eve's conditional entropy, which is given by
\begin{gather}
\sum\limits_{w=1}^{N-1}P(w)S(\vec{L}|E,W=w)\geq  \label{restriction} \\
(N-1)[1-P(w=0)]-NS(A)=:K.  \notag
\end{gather}%
where we have used the fact that $S(\vec{L}|E,W=0)\leq S(\vec{L})=N-1$, and $%
K$ 
denotes the first term in the second line.

Eq. (\ref{restriction}) gives one constraint on $S(\vec{L}|E,W=w)$. There is
also another trivial constraint on $S(\vec{L}|E,W=w)$, which is%
\begin{equation}
S(\vec{L}|E,W=w)\leq S(\vec{L})=N-1.  \label{another restriction}
\end{equation}%
Then the remaining problem is to upper bound Eve's information (or to lower
bound the secret key rate) under the constraints given by Eqs. (\ref%
{restriction}) and (\ref{another restriction}).

\subsection{Lower bound of the secret key rate}

In this part we will give the lower bound of the secret key rate based on
the above analysis. From Eq. (\ref{IAB}) it can be seen that if there is no
bit error, the mutual information between Alice and Bob is
\begin{eqnarray*}
I(\vec{U}_{A} &:&\vec{U}_{B}|\vec{Z})=\sum_{\vec{z}}P(\vec{z})[H(\vec{L}_{%
\vec{z}}|\vec{Z}=\vec{z}) \\
-H(\vec{L}_{\vec{z}}|\vec{Y}_{\vec{z}},\vec{Z} &=&\vec{z})]=\sum_{\vec{z}}P(%
\vec{z})H(\vec{L}_{\vec{z}}|\vec{Z}=\vec{z})
\end{eqnarray*}%
where we have used the fact that if there is no bit error $H(\vec{L}_{\vec{z}%
}|\vec{Y}_{\vec{z}},\vec{Z}=\vec{z})]=0$\textbf{.} After the channel
estimation Alice and Bob can compute this mutual information. It can be seen
that if for all $\vec{z}$s, $\vec{L}_{\vec{z}}$ is evenly distributed, this
mutual information becomes maximized, and $I(\vec{U}_{A}:\vec{U}_{B}|\vec{Z}%
)=\sum_{\vec{z}}P(\vec{z})w(\vec{z})$. For simplification, in the following
we introduce a term $\Delta $ to denote the difference between Bob's actual
information and the maximal information he can get in principle: 
\begin{equation}
\Delta =\sum_{w}P(w)w-I(\vec{U}_{A}:\vec{U}_{B}|\vec{Z}).  \label{delta}
\end{equation}%
Then from Eqs. (\ref{IAB}), (\ref{Eves information}) and (\ref{delta}) we
know that if there is no bit error the secret key rate per $N$ pulses
satisfies%
\begin{equation}
G\geq \frac{1}{N-1}\sum_{w=1}^{N-1}P(w)wS(\vec{L}|E,W=w)-\Delta
\label{lower bound of G}
\end{equation}%
where $P(w)$ and $\Delta $ are known after the channel estimation.

It can be seen that under the constraints of Eqs. (\ref{restriction}) and (%
\ref{another restriction}), the lower bound of the secret key rate given by
Eq. (\ref{lower bound of G}) reaches its minimum when $S(\vec{L}|E,W=w)=0$
for large values of $w$ and $S(\vec{L}|E,W=w)=N-1$ for small values of $w$
while satisfying
Eq. (\ref{restriction}) at the same time. Then the final lower bound on the
secret key rate per $N$ pulses is given by%
\begin{equation}
G\geq \sum_{w=1}^{w_{0}}P(w)w-\Delta  \label{final secret key rate}
\end{equation}%
where the $w_{0}$ is the solution to the equations 
\begin{equation}
\left\{
\begin{array}{c}
\sum\limits_{w=1}^{w_{0}}P(w)(N-1)\leq K \\
\sum\limits_{w=1}^{w_{0}+1}P(w)(N-1)\geq K%
\end{array}%
\right.  \label{solution}
\end{equation}%
where $K$ is defined in the Eq. (\ref{restriction}). This formula
for the key generation rate in Eq.~\eqref{final secret key rate} is
applicable to any $N$ any distribution in Bob's detection
statistics. In the next section, we will consider a particular
distribution as an example.

\section{Secret key rate under the binomial distribution case}

We illustrate how to compute the key generation rate in Eq.~\eqref{final secret key rate} derived in the previous section for a channel
that produces a binomial distribution in Bob's
detection statistics.
We show that in this case the
secret key generation rate per pulse is linearly proportional to the channel
transmission probability (see Eq.~\eqref{key rate per pulse}).

Consider the specific case where Bob's total count obeys the binomial
distribution, $\Delta =0$ and $N\rightarrow \infty $. Under this case
\begin{equation}
P(w)=C_{N-1}^{w}r^{w}(1-r)^{N-w-1}  \label{distribution}
\end{equation}%
where $r$ denotes the detection rate per pulse.

For the $N\rightarrow \infty $ case, the binomial distribution
tends to the Gaussian distribution with 
the same mean and variance as $P(w)$.
For convenience of discussion of the $N\rightarrow \infty $ case, here we
introduce $\lambda $ denote the ratio of $w$ over $N-1$: 
\begin{equation*}
\lambda =\frac{w}{N-1}.
\end{equation*}%
Under the case that $N\rightarrow \infty $, the $\lambda $ can be regarded
as a real number.\ Then instead of 
dealing with the Gaussian approximation of $P(w)$ we can deal with 
an approximation for $P(\lambda )$, which denotes the distribution of $%
\lambda $. The Gaussian approximation of $P(\lambda )$ can be directly given
by
\begin{equation}
P(\lambda )\approx \frac{N-1}{\sqrt{2\pi (N-1)r(1-r)}}\exp [-\frac{%
(N-1)(r-\lambda )^{2}}{2r(1-r)}]  \label{Plamda}
\end{equation}%
which has the same mean and variance as $P(\lambda )$.

The lower bound of the secret key rate given by Eq. (\ref{final secret key
rate}) can be approximated by
\begin{equation}
G\geq (N-1)\int\limits_{0}^{\lambda _{0}}d\lambda P(\lambda )\lambda
\label{key rate for binary}
\end{equation}%
where the $\lambda _{0}=w_{0}/(N-1)$, which can be given by the solution to
the equation%
\begin{equation}
\int\limits_{0}^{\lambda _{0}}d\lambda P(\lambda )=\frac{K}{N-1}
\label{solution for binary}
\end{equation}%
Here Eq. (\ref{solution for binary}) is obtained by putting $\lambda $ into
Eq. (\ref{solution}).

It can be seen that when $N\rightarrow \infty $, we have $P(w=0)=0$ and $%
\frac{N}{N-1}=1$. Then Eq. (\ref{solution for binary}) becomes

\begin{equation}
\int\limits_{0}^{\lambda _{0}}d\lambda P(\lambda )=1-S(A)
\label{final solution}
\end{equation}

To find a solution to Eqs. (\ref{key rate for binary}) and (\ref{final
solution}), we introduce another function $F(\lambda )$: 
\begin{equation}
F(\lambda )=\left\{
\begin{array}{c}
P(\lambda )\text{, while\ }\lambda \leq \lambda _{0} \\
0\text{ while }\lambda >\lambda _{0}%
\end{array}%
\right. .  \label{Flamda}
\end{equation}%
Then it can be seen that when $N\rightarrow \infty $,
\begin{equation*}
\begin{array}{c}
\int\limits_{-\infty }^{\infty }d\lambda F(\lambda )=1-S(A) \\
F(\lambda )=0\text{ for }\lambda \neq r%
\end{array}%
\end{equation*}%
where the expression in the second line can be seen from the fact that $%
\lim_{N\rightarrow \infty }\frac{N}{\sqrt{2\pi Nr(1-r)}}\exp [-\frac{%
N(r-\lambda )^{2}}{2r(1-r)}]=0$ for\textbf{\ }$\lambda \neq r$.

We see that $F(\lambda )$ is actually a delta function satisfying $F(\lambda
)=[1-S(A)]\delta (\lambda -r)$. From Eqs. (\ref{key rate for binary}) and (%
\ref{Flamda}), we can finally get the secret key rate for the binomial
distribution and $N\rightarrow \infty $ case. It is given by
\begin{align}
G& \geq (N-1)\int\limits_{-\infty }^{\infty }d\lambda F(\lambda )\lambda
\label{final key rate} \\
& =(N-1)r[1-S(A)] .  \notag
\end{align}

When given that the amplitude of coherent state is $\alpha $ and the total
transmission probability is $\eta $, then the secret key rate per pulse can
be given by%
\begin{equation}
g\geq \eta |\alpha |^{2}\{1-h[\frac{1}{2}(1-e^{-4|\alpha |^{2}})]\}
\label{key rate per pulse}
\end{equation}%
where we have used Eq. (\ref{SA}), the result that $|\langle -\alpha |\alpha
\rangle |=e^{-4|\alpha |^{2}}$ and the factor $N-1$ is canceled since this
is a key rate per pulse. From this result it can be seen that the lower
bound of the secret key rate per pulse is linearly proportional to the
channel transmission probability. It can be calculated that when $\alpha
=0.338$, the right part in the Eq. (\ref{key rate per pulse}) is maximized
and is given by $0.0357\eta $. Note that our result is consistent with the
upper bounds on the key rate 
given in Refs. \cite{Upper bound,Sequential attack}.
From Ref. \cite{Lo2005}, one can easily find that the key generation
rate per tranmistted pulse of BB84 in the noiseless case scales at a
higher rate than that of DPSQKD. However, the overall secret key
generation rate is also determined by the modulation rate, and we
remark that it is possible that DPSQKD can outperform BB84 in the
modulation rate, since DPSQKD only requires one binary phase
modulation at Alice's side, while BB84 requires a quaternary
modulator at Alice's side and a binary or quaternary modulator at
Bob's sides \cite{Fred} (the binary modulation is easier to realize
than the quaternary one.).

\section{Conclusion}

We prove the security for DPSQKD with a weak-coherent light source
against collective attacks in the noiseless case. The only
assumption we employ are that the quantum theory is true, the device
is trusted and the key size is infinite. The key point that
guarantees this scheme to be secure is that Eve's state is
independent of the positions of Bob's detections, so that after the
post-selection, in which Alice and Bob discard the data that Bob
did not receive a signal for, 
Bob knows Alice's sifted data better than Eve does. In addition, we consider
a specific case where the total number of Bob's count obeys the binomial
distribution. 
In this case, we derive the lower bound of the secret key rate per pulse and
it is linearly proportional to the channel transmission probability. This
result definitely suggests that DPSQKD has a high potential for high speed
communication, since it is easy to engineer DPSQKD to operate at a high
modulation rate. Although we have only proved the security of DPSQKD for the
noiseless case, we hope that our work can offer some insights into the
security of DPSQKD and may serve as a stepping stone for proving the
security for the noisy case.

\textbf{Acknowledge: }Special thanks are given to Norbert
L\"{u}tkenhaus and Matthias Heid for their innumerable discussions
on security proof. Thanks are also given to H.-K. Lo for fruitful
discussions. This work is supported in part by the National
Fundamental Research Program of China under Grant No 2006CB921900,
National Natural Science Foundation of China under Grants No.
60537020 and 60621064, Innovation Fund of the University of Science
and Technology of China under Grant No. KD2006005, the Knowledge
Innovation Project of the Chinese Academy of Sciences (CAS), the
Research Grants Council of Hong Kong under Grant No. HKU 701007P,
and the Natural Sciences and Engineering Research Council of Canada.

\appendix

\section{Super subadditivity for the multi-system case}

\begin{theorem}
\label{thm-subadd} Suppose $\Theta _{n}=\{A_{1},\ldots,A_{n}\}$ is a
collection of $n$ systems and $\Xi _{m}^{n}$ is a collection of all possible
$m$ selections from $n$ cases.\ Then it is always true for arbitrary $n$ and
$m\leq n$ that
\begin{equation}  \label{eqn-subadd1}
\sum\limits_{\vec{i}\in \Xi
_{m}^{n}}S(A_{i_{1}},A_{i_{2}},\ldots,A_{i_{m}}|E)\geq C_{n-1}^{m-1}
S(A_{1},A_{2},\ldots,A_{n}|E)
\end{equation}%
where $\vec{i}=(i_1,i_2,\ldots,i_m)$, $E$ is another system, and $C_{n}^{m}=%
\frac{n!}{m!(n-m)!}$.
\end{theorem}

This theorem is a generalization of the following lemma for subadditivity of
Von Neuemann entropy (see, e.g., \cite{Nielson}).

\begin{lemma}
\label{lemma1} For three quantum systems $A_1$, $A_2$, and $E$, $%
S(A_{1}|E)+S(A_{2}|E)\geq S(A_{1},A_{2}|E)$.
\end{lemma}

\begin{proof}[Proof of Theorem~\ref{thm-subadd}]
We will use induction to prove the theorem.
First, notice that for $m=n$, this theorem is always true.
%
In the following we will prove that for $n$ and $m=1$ this theorem
holds, which will be called first proof (FP) in the following.

Then we prove if for $n$ and $m$ and $m-1$ ($m>1$) the above theorem is
correct, then for $n+1$ and $m$ it is also correct, which will be called
second proof (SP).

Thirdly, we will prove that if for $n-1$ and $m=n-2$ ($n>2$) the above
theorem is true then for $n$ and $m+1$ it is also valid, which will be
called third proof (TP).

Since for $n=2$ and $m=1$, the above theorem is true, then it is also true for $n=3$ and $m=2$ by applying the TP.

This theorem is true for $n=3$ and $m=2$ and $n=3$ and $m=1$. Then by applying the SP,
we know that it is also valid for $n=4$ and $m=2$ and consequently for all
of $n$ and $m=2$.

After we continuously do such induction we can see that if FP, SP and TP is
correct then for all $n\geq m$ the above theorem holds.

Now we prove the FP first.
Observe that for arbitrary $n$ and $m=1$, Eq.~\eqref{eqn-subadd1} holds by repeated applications of Lemma~\ref{lemma1}, i.e.
\begin{equation}
\label{eqn-subadd2}
\sum_{i=1}^{n}
S(A_i|E)\geq
S(A_{1},A_{2},\ldots,A_{n}|E)
\end{equation}%
holds.

Now we will prove the SP, which is claims that if for $n$ and $m$
and $m-1$ the above theorem holds, then for $n+1$ and $m$ it also
holds.

We assume that $\Theta
_{n+1}=\{A_{1},...,A_{n},A_{n+1}\}=\{A_{1},...,A_{n},B\} $. Here we use $B$ to
denote $A_{n+1}$.

In the following we use $\Xi _{m}^{n}$ to denote the collection of all
possible $m$ selections in $n$ samples.

First we have %
\begin{eqnarray}
T &=&\sum\limits_{\vec{i}\in \Xi
_{m}^{n+1}}S(A_{i_{1}},A_{i_{2}},...,A_{i_{m}}|E)  \notag \\
&=&\sum\limits_{\vec{i}\in \Xi _{m}^{n}}S(A_{i_{1}},A_{i_{2}},...,A_{i_{m}}|E)
\notag \\
&&+\sum\limits_{\vec{i}\in \Xi
_{m-1}^{n}}S(A_{i_{1}},A_{i_{2}},...,A_{i_{m-1}},B|E)  \notag \\
&=&C+D  \label{T}
\end{eqnarray}%
where $\vec{i}=(i_{1},i_{2},...,i_{m})$ denotes a possible $m$ selection in $%
n$ cases and we introduced $T$, $C$ and $D$ to denote the terms given in the
first, second and third line respectively. If the theorem is valid for $n$
and $m$, then we have
\begin{equation}
C\geq \frac{mC_{n}^{m}}{n}S(A_{1},A_{2},...,A_{n}|E)  .\label{C}
\end{equation}%
Now, $D$ can also be given by%
\begin{eqnarray}
D &=&\sum\limits_{\vec{i}\in \Xi
_{m-1}^{n}}S(A_{i_{1}},A_{i_{2}},...,A_{i_{m-1}}|BE)+C_{n}^{m-1}S(B|E)  \notag
\\
&=&D_{1}+C_{n}^{m-1}S(B|E)  \label{D}
\end{eqnarray}%
where we have applied the fact that $S(A|EB)+S(B|E)=S(AB|E)$ and $%
\sum\limits_{\vec{i}\in \Xi _{m-1}^{n}}=C_{n}^{m-1}$. Also, if the above
theorem is correct for $n$ and $m-1$, we have
\begin{equation}
D_{1}\geq \frac{(m-1)C_{n}^{m-1}}{n}S(A_{1},A_{2},...,A_{n}|BE)  .\label{D1}
\end{equation}%
Here,
\begin{equation}
C_{n}^{m}=\frac{n!}{m!(n-m)!}  .\label{Cnm}
\end{equation}%
Then if we put Eqs. (\ref{C}), (\ref{D}), (\ref{D1}) and (\ref{Cnm}) into
Eq. (\ref{T}) we can immediately get
\begin{eqnarray}
T &\geq &\frac{n!(n-m+1)m}{nm!(n-m+1)!}%
[S(A_{_{1},}A_{_{2},}...,A_{n}|E)+S(B|E)]  \label{second t} \\
&&+\frac{n!m(m-1)}{nm!(n-m+1)!}[S(A_{1},A_{2},...,A_{n}|BE)+S(B|E)]  .\notag
\end{eqnarray}

Since the above theorem is correct for $n=2$ and $m=1$, we have
\begin{equation}
S(A_{_{1},}A_{_{2},}...,A_{n}|E)+S(B|E)\geq S(A_{_{1},}A_{_{2},}...,A_{n},B|E).
\label{inequal}
\end{equation}

Then if we put Eq. (\ref{inequal}) into Eq. (\ref{second t}) and apply the
results that $S(A|EB)+S(B|E)=S(AB|E)$ and
\begin{equation*}
\frac{n!m(m-1)}{nm!(n-m+1)!}=\frac{mC_{n+1}^{m}}{n+1},
\end{equation*}
we can obtain%
\begin{equation*}
T\geq \frac{mC_{n+1}^{m}}{n+1}S(A_{1},A_{2},...,A_{n},B|E)
\end{equation*}%
which says that if for $n$ and $m$ and $m-1$, the above theorem is
correct then for $n+1$ and $m$ it is also correct. Now the SP is proved.

Now, we will prove the TP, that is if for $n-1$ and $m=n-2$ ($n>2$) the
above theorem is true then for $n$ and $m+1$ it is also true.

We assume that $\Theta _{n}=\{A_{1},...,A_{n-1},B\}$. Here we use $B$ to
denote $A_{n}$. Then we have that%
\begin{eqnarray}
S &=&\sum\limits_{\vec{i}\in \Xi
_{m+1}^{n}}S(A_{i_{1},}A_{i_{2},}...,A_{i_{m}},A_{i_{m+1}}|E)  \notag \\
&=&S(A_{1,}A_{2},...,A_{n-1}|E)  \notag \\
&&+\sum\limits_{\vec{i}\in \Xi
_{m}^{n-1}}S(A_{i_{1}},A_{i_{2}},...,A_{i_{m}},B|E)  \notag \\
&=&S_{1}+S_{2}  \label{S}
\end{eqnarray}%
where the $S$, $S_{1}$ and $S_{2}$ are introduced to denote the expression
in the first, second and third line respectively, and in the second line we
have used the requirement that $m=n-2$.

Since $S(A|EB)+S(B|E)=S(AB|E)$, and $\sum\limits_{\vec{i}\in \Xi
_{m}^{n-1}}=C_{n-1}^{m}$, 
\begin{eqnarray}
S_{2} &=&\sum\limits_{\vec{i}\in \Xi
_{m}^{n-1}}S(A_{i_{1},}A_{i_{2},}...,A_{i_{m}}|BE)+C_{n-1}^{m}S(B|E)  \notag
\\
&\geq &\frac{mC_{n-1}^{m}}{n-1}S(A_{1,}A_{1,}...,A_{n-1}|BE)+C_{n-1}^{m}S(B|E)
\notag \\
&=&m[S(A_{1,}A_{1,}...,A_{n-1}|BE)+S(B|E)]+S(B|E)  \notag \\
&=&mS(A_{1,}A_{1,}...,A_{n-1},B|E)+S(B|E)  \label{S2}
\end{eqnarray}%
where we have applied the assumption that for $n-1$ and $m$ the above theorem is
correct and the fact that $C_{n-1}^{m}=n-1$ for $m=n-2$.

Now we put Eq. (\ref{S2}) in to Eq. (\ref{S}) we can obtain that
\begin{eqnarray*}
S &\geq &mS(A_{1,}A_{1,}...,A_{n-1},B|E) \\
&&+S(A_{1,}A_{2},...,A_{n-1}|E)+S(B|E) \\
&\geq &(m+1)S(A_{1,}A_{1,}...,A_{n-1},B|E) \\
&=&\frac{(m+1)C_{n}^{m+1}}{n}S(A_{1},A_{2},...,A_{n-1},A_{n}|E)
\end{eqnarray*}%
where in the third line we have used the subadditivity for $n=2$ and $m=1$
case. Now the TP is proved.

Since the assumptions FP, SP and TP hold and the initial conditions
are satisfied, it is proved that for all $n\geq m$ the above theorem
holds.
\end{proof}


\begin{thebibliography}{99}
\bibitem{explanation of collective} Generally, Eve's attack can be separated
into three classes: individual attacks, collective attacks and coherent
attacks \cite{Review,Renner}. In individual attacks, Eve attacks each signal
system separately with the same method and later measures her quantum state
right after the sifting step. In collective attacks, Eve attacks each signal
system separately with the same method but can perform arbitrary measurement
after all of the steps, including the reconciliation, privacy amplification
and the encryption. Coherent attacks are the most general attacks, where Eve
can attack all of the signals together and perform the measurement at the
end of the protocol.
\bibitem{ideal} Here we mean that all of Bob's detectors are the same with
perfect efficiency and there is no dead time associated with them.
\end{thebibliography}
\end{document}